\documentclass{article}
\usepackage{arxiv}

\usepackage{subfig}
\usepackage{todonotes}
\usepackage{listings}
\usepackage{graphicx}
\usepackage{natbib}
\usepackage{amsmath}
\usepackage{footmisc}
\usepackage{hyperref}
\usepackage{booktabs}
\usepackage[utf8]{inputenc} 
\usepackage[T1]{fontenc}    
\usepackage{url}            

\lstdefinelanguage{Jayvee}{
  morekeywords=[1]{block, oftype, pipeline, composite, blocktype, property, input, output, constraint, valuetype, on, range},
  morekeywords=[2]{true, false, text, decimal, integer, None, Sheet, value},
  morekeywords=[3]{},
  sensitive,
  morecomment=[s]{/*}{*/},
  morecomment=[l]//,
  morecomment=[s]{/**}{*/}, 
  morestring=[b]',
  morestring=[b]"
}[keywords, comments, strings]

\title{Is spreadsheet syntax better than numeric indexing for cell selection?}

\newif\ifuniqueAffiliation
\uniqueAffiliationtrue

\ifuniqueAffiliation 
\author{ \href{https://orcid.org/0000-0002-4236-2689}{\includegraphics[scale=0.06]{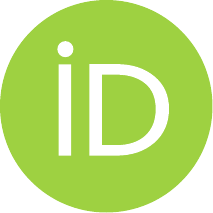}\hspace{1mm}Philip~Heltweg} \\
	Friedrich-Alexander-Universität Erlangen-Nürnberg\\
	Erlangen, Germany \\
	\texttt{philip@heltweg.org} \\
        \And
        \href{https://orcid.org/0000-0002-8139-5600}{\includegraphics[scale=0.06]{orcid.pdf}\hspace{1mm}Dirk~Riehle} \\
	Friedrich-Alexander-Universität Erlangen-Nürnberg\\
	Erlangen, Germany \\
	\texttt{dirk@riehle.org} \\
	\And
	\href{https://orcid.org/0000-0001-9060-7938}{\includegraphics[scale=0.06]{orcid.pdf}\hspace{1mm}Georg-Daniel~Schwarz} \\
	Friedrich-Alexander-Universität Erlangen-Nürnberg\\
	Erlangen, Germany \\
	\texttt{georg.schwarz@fau.de} \\
}
\else
\usepackage{authblk}

\newbox{\orcid}\sbox{\orcid}{\includegraphics[scale=0.06]{orcid.pdf}} 
\author[1]{%
	\href{https://orcid.org/0000-0000-0000-0000}{\usebox{\orcid}\hspace{1mm}David S.~Hippocampus\thanks{\texttt{hippo@cs.cranberry-lemon.edu}}}%
}
\author[1,2]{%
	\href{https://orcid.org/0000-0000-0000-0000}{\usebox{\orcid}\hspace{1mm}Elias D.~Striatum\thanks{\texttt{stariate@ee.mount-sheikh.edu}}}%
}
\affil[1]{Department of Computer Science, Cranberry-Lemon University, Pittsburgh, PA 15213}
\affil[2]{Department of Electrical Engineering, Mount-Sheikh University, Santa Narimana, Levand}
\fi

\renewcommand{\headeright}{Preprint}
\renewcommand{\undertitle}{Preprint}

\hypersetup{
    pdftitle={Is spreadsheet syntax better than numeric indexing for cell selection?},
    pdfauthor={Philip~Heltweg, Dirk~Riehle, Georg-Daniel~Schwarz},
}

\begin{document}
\newpage
\maketitle

\begin{abstract}
Selecting a subset of cells is a common task in data engineering, for example, to remove errors or select only specific parts of a table. Multiple approaches to express this selection exist. One option is numeric indexing, commonly found in general programming languages, where a tuple of numbers identifies the cell. Alternatively, the separate dimensions can be referred to using different enumeration schemes like "A1" for the first cell, commonly found in software such as spreadsheet systems.

In a large-scale controlled experiment with student participants as proxy for data practitioners, we compare the two options with respect to speed and correctness of reading and writing code.

The results show that, when reading code, participants make less mistakes using spreadsheet-style syntax. Additionally, when writing code, they make fewer mistakes and are faster when using spreadsheet syntax compared to numeric syntax.

From this, a domain-specific syntax, such as spreadsheet syntax for data engineering, appears to be a promising alternative to explore in future tools to support practitioners without a software engineering background.
\end{abstract}


\section{Introduction}
\label{sec:introduction}

A common challenge in data engineering is working with unstructured, two-dimensional data as it can be found in CSV files or spreadsheet software. Especially data sets based on exports from spreadsheets made for human readers have to be wrangled without the easy-to-use format that would allow for the selection of cells by column names or other structured tools.

In these cases, data is organized for human readers to consume rather than machines. Often, values are distributed in a 2D data structure to place them in a 2D space when displayed as part of a sheet. While these data sets are technically machine-readable, they have to undergo extensive data engineering work before they are available in a format that is easily importable into, e.g., a Pandas dataframe.

General-purpose programming languages (GPLs) can be used to manipulate this data, e.g., to select a subsection of it or remove errors. However, the syntax used in most GPLs has its origin in the numeric and often zero-indexed access of array structures. To select cells using this syntax, programmers refer to columns and rows by numbers or numeric ranges with the distinction between both being mostly by position. For example, Pandas/Python cell selection is performed by the axis of the underlying dataframe so that \lstinline{df.iloc[0, 1]} selects the cell in the first row (on index 0) and second column (on index 1). 

While this syntax is familiar to professional software engineers, practitioners from adjacent fields that also work with two-dimensional (2D) data often use spreadsheet software to manipulate cells visually. Popular tools such as Microsoft Excel or Google Sheets show numbers as the reference to rows and characters to refer to columns, thereby syntactically separating the indexing dimensions. Therefore, the equivalent reference string to the Python/Pandas example above would be \lstinline{B1}. In this syntax, columns are expressed as characters (\lstinline{B} for the second column) while references to rows stay numeric, but their index starts more naturally at one instead of zero.

With these different syntactical approaches, data practitioners with a background in using spreadsheet software to manipulate and clean their datasets can struggle when reading or writing code using numeric indexing. Mistakenly switching the order of row and column references or forgetting about zero-indexing can lead to crashes or subtle errors in the resulting datasets.

As an alternative, domain-specific languages (DSLs) have been shown to be a potential middle ground between GPLs and visual tools, enabling domain experts to efficiently contribute in a wide range of domains outside of data engineering \citep{Kosar2018-ck, Johanson2017-sb}, as well as when building data pipelines \cite{Heltweg2025-uc}. They do so by re-using concepts and conventions from the domain they cover. It stands to reason that addressing cells in unstructured 2D data sets could be easier for data practitioners using spreadsheet syntax rather than numeric indexing.

In this study, we conduct a controlled experiment to test this hypothesis and provide a basis for further research. Based on quantitative data from a large group of student participants, we compare the use of spreadsheet-style syntax in a DSL for data engineering with the numeric syntax of Pandas, an industry-standard data engineering library for Python.

Our goal is to answer the following research questions:

\textbf{Research Question 1:} Does spreadsheet-style syntax have an effect on \textit{bottom-up program comprehension of} cell selection in unstructured 2D data by data practitioners compared to numeric syntax...

\textbf{a:} regarding speed?

\textbf{b:} regarding correctness?

\textbf{Research Question 2:} Does spreadsheet-style syntax have an effect on \textit{code creation} for cell selection in unstructured 2D data by data practitioners compared to numeric syntax...

\textbf{a:} regarding speed?

\textbf{b:} regarding correctness?

With this study, we contribute:

\begin{enumerate}
    \item The results of hypothesis tests in a controlled experiment on the effects of using spreadsheet-style syntax instead of numeric syntax for cell selection, providing a foundation for future studies towards the ideal syntax for DSLs in data engineering.
    \item A detailed description and accompanying code release of an experiment instrument to run controlled experiments for cell selection that enables reproduction and re-execution of the experiment with data professionals.
\end{enumerate}

This article is organized as follows:

First, we provide a short overview of related studies in \autoref{sec:related-work}, then we present the research design in \autoref{sec:research-design}, followed by the results in \autoref{sec:results}. We provide additional context and insights in the discussion in \autoref{sec:discussion}. Limitations to the results are presented in \autoref{sec:limitations} before a summary of the insights and future work in \autoref{sec:conclusions}.
\section{Related Work}
\label{sec:related-work}
Empirical studies of programming language design, such as different syntax, with users are generally rare, even though they can lead to a deeper understanding that can not otherwise be achieved \cite{Buse2011-ra}.

The relatively low amount of controlled experiments is a challenge in the wider field of software engineering research as well \cite{Ko2015-ti, Vegas2016-dr}.

The benefits of using domain concepts with DSLs have been investigated in a series of controlled experiments and replication studies by Kosar et al. They compare DSLs with GPLs and domain-appropriate libraries in a variety of domains from GUI programming to feature diagrams \cite{Kosar2010-cp, Kosar2012-ap, Kosar2018-ck}. The domain-specific concepts allow participants to work more accurate and efficient, both with and without IDE support.

However, based on their experience they point out that the evaluation of DSLs and their features is domain-specific and should be done for each domain. To the best of our knowledge, our study is the first contribution to this effort for cell selection in data engineering.

Hoisl et al. \cite{Hoisl2014-jy} compare notation for scenario-based testing (two text-based and one diagrammatic) and find the more natural-language based one to outperform others, including a structured language. Their work is an indication that task outcomes can be improved by syntax that is more familiar to the participants that we test in a separate domain.

How deeply domain concepts should be part of a DSL was previously studied by Häser et al. \cite{Haser2016-at} for behavior driven development who compared a simple DSL with a DSL that was enriched with domain concepts. They find that participants can complete tasks significantly quicker with the DSL that includes domain concepts without an effect on quality.

Recently, Klanten et al. \cite{Klanten2024-pf} have evaluated a similar approach to using domain-specific syntax (a domain-specific syntax for type inference rules compared to an implementation in Java) with positive effects for speed and correctness.

Similar to these studies, our experiments contributes additional data towards the study of optimal DSL features in a specific domain, in this case with regards to cell selection syntax in data engineering.
\section{Research Design}
\label{sec:research-design}
We gathered quantitative data using a controlled experiment with human participants. To do so, we first defined a plan for the experiment that we refined iteratively with pilot experiments with other researchers. Then, we executed the experiment in-person during multiple sessions on one day. Finally, we analyzed the data using standard statistical methods.

The following structure is adapted from the proposed guidelines for reporting controlled experiments as suggested in Wohlin et al. \cite{Wohlin2012-ze}, adapted from Jedlitschka and Pfahl \cite{Jedlitschka2005-mu}.

\subsection{Problem Statement}
Working with otherwise unstructured 2D data is a common task in data engineering, for example when handling data from CSV files. When the data is formatted like a table with a clear header row, column names can be used to select subsets. However, often header data is missing or complex, multi-line headers based on exports from human-readable sheets make simple indexing by column names impossible.
In those cases, subsets of data need to be selected first to be extracted, deleted, or transformed in follow-up steps. Different ways to manipulate these data structures exist:

|\begin{enumerate}
    \item General-purpose programming languages with libraries, such as Python and Pandas, using numeric indexing like iloc, e.g., \lstinline{df.iloc[1:5, 2:4]}
    \item Domain-specific spreadsheet software like Microsoft Excel or Google Docs and their syntax, e.g., \lstinline{A1:B10}
\end{enumerate}

While the numeric indexing might be familiar to software developers and especially users of Python/Pandas, spreadsheet-software is widely used by subject-matter experts and data practitioners in data engineering.
For a domain-specific language for data engineering that aims to enable subject-matter experts to contribute, it is unclear whether using numeric indexing or spreadsheet-software formula syntax is the better choice to select subsets of 2D data structures.

\subsection{Research Objectives}
We follow the Goal/Question/Metric template \citep{Wohlin2012-ze, Basili1988-al} to define the research objective:

\begin{itemize}
    \item Analyze \textit{two cell selection syntaxes}
    \item for the purpose of their effect on \textit{bottom-up program comprehension and code creation for cell selection}
    \item with respect to \textit{speed and correctness}
    \item from the point of \textit{view of researchers}
    \item in the context of \textit{a university course with masters level students learning data science (as proxies for data practitioners)}
\end{itemize}

\subsection{Context}
The context of the experiment is a master's level course teaching advanced methods of data engineering. Students were largely from master’s degrees in artificial intelligence and data science. Controlled experiments with students allow for the initial evaluation of hypotheses that can be extended by experiments with practitioners in future research if they are a conscious choice as a representative of a population \cite{Falessi2018-so, Tichy2000-nq}.

We consider this cohort of students as an appropriate proxy for our target population of data practitioners that are used to working with data as part of their job but do not have a professional programming background. Similar to them, students from artificial intelligence and data science have heard lectures on statistics and theoretical algorithms but lack experience in professional software development. From previous work with a similar cohort of students, we know that they also use spreadsheet software to work with data in addition to writing scripts \cite{Heltweg2025-uc}.

The participants learned an open-source domain-specific language called Jayvee \cite{Heltweg2025-uc} using spreadsheet-style cell selection syntax during the course. They did so by listening to introductory lectures and completing data engineering exercises while implementing a self-directed data science project in Python.

\subsection{Experimental Design}

\subsubsection{Goals, Hypotheses, Parameters, and Variables}

We derived two goals from the research objectives.

Goal 1: Understand if the use of spreadsheet-software cell selection syntax has an effect on bottom-up program comprehension ($pc$) compared to numeric indexing in regard to

\begin{enumerate}
    \renewcommand{\theenumi}{\alph{enumi}}
    \item speed
    \item correctness
\end{enumerate}

Goal 2: Understand if the use of spreadsheet-software cell selection syntax has an effect on code creation ($cc$) compared to numeric indexing in regard to

\begin{enumerate}
    \renewcommand{\theenumi}{\alph{enumi}}
    \item speed
    \item correctness
\end{enumerate}

During the experiment, we defined variables and controlled the following parameters.

\textbf{Parameters}

\begin{enumerate}
    \item The tasks, based on two real open data sets. We selected two different, real data sets that were understandable without any special domain knowledge and slightly adapted them by removing rows with empty values, selecting a subset of 10 by 10 cells and randomizing the order of rows.
    \item The students, from a master's level university course on data engineering.
    \item The programming environment, an in-person experiment on a web-based experiment tool. The experiment tool did not provide syntax highlighting or auto completion of any sort and ensured that individual editor choice had no influence on the results.
    \item Available help, only allowing standard documentation. We asked experiment participants to open the official documentation of both treatments before the start of the experiment and to not use the internet in any other way. One researcher ensured that participants did not leave the provided experiment environment at all times.
\end{enumerate}

\textbf{Independent variables}

\begin{enumerate}
    \item The cell selection syntax used, either a DSL with \textit{Spreadsheet} syntax or Python with Pandas and it's iloc cell selection using \textit{Numeric} syntax
\end{enumerate}

\textbf{Dependent variables}

\begin{enumerate}
    \item The average \textit{time} to task completion, in seconds from the moment a participant started a task until they decided to submit their solution.
    \item The average correctness of solution defined as defined by the Jaccard index \cite{Jaccard1901-yx}: $J(C,S) = \frac{|C \cap S|}{|C \cup S|}$ where $C$ is the set of cells that should be selected and $S$ the set of cells that are selected.
\end{enumerate}

From the goals and based on the variables we measure, we defined the following hypotheses to test.

For goal 1a, the null hypothesis is \textit{"Using spreadsheet-software cell selection syntax has no effect on the speed of bottom-up program comprehension compared to numeric indexing"}, more formally:

\begin{equation}
    \begin{aligned}
        H_{0, 1a}: time_{pc}(Spreadsheet) = time_{pc}(Numeric) \\
        H_{1, 1a}: time_{pc}(Spreadsheet) \neq time_{pc}(Numeric)
    \end{aligned}
\end{equation}

For goal 1b, the null hypothesis is \textit{"Using spreadsheet-software cell selection syntax has no effect on the correctness of bottom-up program comprehension compared to numeric indexing"}, more formally:

\begin{equation}
    \begin{aligned}
        H_{0, 1b}: correctness_{pc}(Spreadsheet) = correctness_{pc}(Numeric) \\
        H_{1, 1b}: correctness_{pc}(Spreadsheet) \neq correctness_{pc}(Numeric)
    \end{aligned}
\end{equation}

For goal 2a, the null hypothesis is \textit{"Using spreadsheet-software cell selection syntax has no effect on the speed of code creation compared to numeric indexing"}, more formally:

\begin{equation}
    \begin{aligned}
        H_{0, 2a}: time_{cc}(Spreadsheet) = time_{cc}(Numeric) \\
        H_{1, 2a}: time_{cc}(Spreadsheet) \neq time_{cc}(Numeric)
    \end{aligned}
\end{equation}

For goal 2b, the null hypothesis is \textit{"Using spreadsheet-software cell selection syntax has no effect on the correctness of code creation compared to numeric indexing"}, more formally:

\begin{equation}
    \begin{aligned}
        H_{0, 2b}: correctness_{cc}(Spreadsheet) = correctness_{cc}(Numeric) \\
        H_{1, 2b}: correctness_{cc}(Spreadsheet) \neq correctness_{cc}(Numeric)
    \end{aligned}
\end{equation}

We chose two-tailed hypotheses because we had no prior knowledge about the effect direction that we expected.

\subsubsection{Experiment Design}
We chose a crossover design for our experiment, a within-subjects design in which each participant is assigned to every treatment. We chose a crossover design because students can have different previous experiences which could lead to challenges when measuring differences between participants groups instead of differences to the participants average \cite{Vegas2016-dr}. In addition, crossover designs are commonly used in software engineering research and well understood \citep{Wyrich2023-ji}.

However, because each participant is assigned to all treatments, crossover designs can introduce carryover effects in which experience from previous tasks influences the completion of future tasks. To reduce this effect, we assigned participants to two different sequences and introduced an initial non-tracked task in pseudocode that allowed them to get familiar with the experiment tool instead of having to learn it during the first real tasks.

Participants were randomly assigned to two sequences AB (first spreadsheet syntax, then numeric indexing) and BA (first numeric, then spreadsheet). To study the effects on both program comprehension and code creation, we ran two different sets of tasks. A short description of the goal of each task is shown in \autoref{tab:experiment-task-description}. For both code creation and program understanding, our goal was to offer one task that includes full rows, one that includes full columns, and two tasks that handle different subsets of cells.

In one session, each participant completed both sets of tasks with the following task order: For code creation, task 1 to 4, see \autoref{tab:experiment-design-code creation}. For program comprehension, task 5 to 8, see \autoref{tab:experiment-design-program-comprehension}.

\begin{table}[h]
    \centering
    \caption{Task descriptions.}
    \label{tab:experiment-task-description}
    \begin{tabular}{cl}
    \toprule
    Task & Description \\
    \midrule
    1 & Understand code that selects complete rows\\
    2 & Understand code that selects connected cells,\\
      & no complete rows/columns\\
    3 & Understand code that selects complete columns\\
    4 & Understand code that selects connected cells,\\
      & no complete rows/columns\\
    5 & Write code to select complete rows\\
    6 & Write code to select connected cells,\\
      & no complete rows/columns\\
    7 & Write code to select complete columns\\
    8 & Write code to select connected cells,\\
      & no complete rows/columns\\
    \bottomrule
    \end{tabular}
\end{table}

\begin{table}[h]
    \centering
    \caption{Sequences and intervention assignment for code creation.}
    \label{tab:experiment-design-code creation}
    \begin{tabular}{lllll}
    \toprule
    Sequence & \multicolumn{4}{c}{Period}  \\
     & Task 1  & Task 2 & Task 3 & Task 4  \\
    \midrule
    AB & Spreadsheet & Numeric & Spreadsheet & Numeric\\
    BA & Numeric & Spreadsheet & Numeric & Spreadsheet\\
    \bottomrule
    \end{tabular}
\end{table}

\begin{table}[h]
    \centering
    \caption{Sequences and intervention assignment for program comprehension.}
    \label{tab:experiment-design-program-comprehension}
    \begin{tabular}{lllll}
    \toprule
    Sequence & \multicolumn{4}{c}{Period}  \\
     & Task 5  & Task 6 & Task 7 & Task 8 \\
    \midrule
    AB & Spreadsheet & Numeric & Spreadsheet & Numeric\\
    BA & Numeric & Spreadsheet & Numeric & Spreadsheet\\
    \bottomrule
    \end{tabular}
\end{table}

To reduce learning effects between the sets of tasks, the dataset used for program comprehension was different from the one for code creation. In any case, the datasets were real-world open data sets, slightly edited to remove header information and empty values, randomize the order of columns, and standardize them to a 10 by 10 grid.

Participants were assigned randomly to a sequence using JavaScript's built-in \lstinline{Math.random} method when opening the experiment tool.

\subsection{Participants}
Participants for the experiment were selected by convenience sampling from students of a master's level course for advanced methods of data engineering, mostly from master’s degrees in artificial intelligence and data science with some students from related degrees such as information systems. Students were familiar with both treatment syntaxes by previous participation the course.

Approval by an ethics committee is not required by our institution and not standard for these kinds of studies as they do not carry personal risk or undue burden. However, we shared an informed consent handout with participants detailing experiment goals, data handling and process by email. Immediately before the experiment started, we again shared the same informed consent handout, allowed time for questions and asked for an explicit opt-in while making it clear that not participating at this point would have no negative consequences.

The handout made explicit, that the students' performance in the experiment tasks had no effect on their course grade, however, we incentivized participation by rewarding points for the course grade for participation, independent of their performance or whether they opted into the use of their data for research purposes. Opting in to allow the use of their data was purely voluntary and had no effect on the grades of students.

We asked for permission to use the data before starting the experiment to make the opt-in independent of the performance during the tasks.

\subsection{Objects}
The experiment was carried out using a web-based experiment tool with a small CSV data set displayed as a table without header.

There were two types of tasks: code creation and code understanding. Before every task, a not-tracked example task in pseudocode allowed the participants to learn how the experiment tool worked and what would be expected of them. Additionally, the experimenters demonstrated the different task types at the start of the experiment.

First, for code creation, participants were shown the data on the left side of the tool. For every tasks, a different subset of the data was highlighted in blue. On the right side, participants were shown a short program excerpt (a snippet from the experiment tasks in the DSL is shown in \autoref{listing:example-code-creation}, numeric syntax tasks used equivalent Python/Pandas code) and asked to complete a code block selecting the highlighted cells, either using numeric indexing with the iloc API in Python/Pandas or spreadsheet-software syntax in the DSL.

\begin{lstlisting}[language=Jayvee, caption={A DSL code snippet participants had to complete for a code creation task. An input field after the range keyword allows for spreadsheet-style syntax to select cells.\label{listing:example-code-creation}}]
    // Other blocks and pipeline definition... 
    
    block DataSelector oftype CellRangeSelector {
      select: range                   ;
    }
\end{lstlisting}

For numeric syntax using Pandas/Python, code creation tasks where based on the participants selecting cells based on position using the \lstinline{iloc} API. The surrounding Python/Pandas code was provided during the experiment tasks so that participants only needed to use the numeric syntax inside the \lstinline{iloc} call to complete the selection.

Using \lstinline{iloc}, participants can select a subset of a Pandas dataframe using a variety of ways such as integers, arrays of integers or slice objects to refer to cell positions. During the lectures in preparation for the experiment the use of mainly integers or slice objects was highlighted (for example \lstinline{df.iloc[0, 1]} or \lstinline{df.iloc[1:3]}).

The DSL used in the experiment is based on connecting small blocks of computation using pipes. These blocks have to be configured by the user, with a block named \lstinline{CellRangeSelector} allowing the selection of a subset of cells from 2D data. The syntax used to select a range of cells using this block aligns with common spreadsheet programs where ranges are described from the starting cell to a final cell (for example, \lstinline{A1:B2} refers to the range from cell \lstinline{A1} to \lstinline{B2}).

Cells are referred to either by a character for the column, followed by a one-indexed number for the row (for example \lstinline{B2} for the second column and second row). Additionally, either the column or the row reference can be replaced by a \lstinline{*} to indicate the last cell in that row or column, allowing a syntax like \lstinline{A1:B*} to select all cells in the first two columns of a data set.

In the same way as for the numeric syntax, all custom code for the DSL was provided and did not have to be remembered by the participants. They only needed to complete the \lstinline{select} property of a block using cell selection syntax.

\begin{figure}[h]
  \centering
  \includegraphics[width=\linewidth]{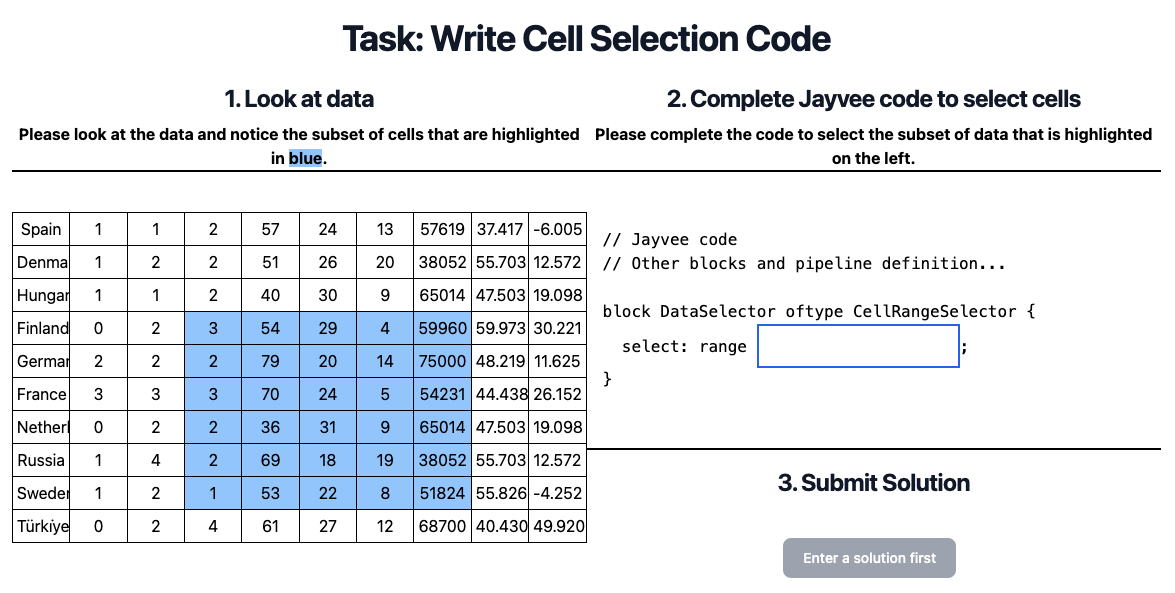}
  \caption{Code creation task in the experiment tool.}
  \label{fig:experiment-tool-cc}
\end{figure}

An example screenshot of the whole task screen in the experiment tool is shown in \autoref{fig:experiment-tool-cc}.

The second type of task was aimed at testing bottom-up program comprehension. Participants were shown code that selects a subset of cells from a 2D data structure in either the DSL or Python/Pandas (using numeric indexing, an example code snippet in Python from the experiment is shown in \autoref{listing:example-program-comprehension}, equivalent DSL code was shown for spreadsheet syntax).

On the right, participants were shown the actual data in the same view as during the code creation tasks. They could highlight cells in blue by either clicking them or dragging the mouse and were asked to highlight the cells that will be selected with the code shown.

\begin{lstlisting}[language=Python, caption={Python code excerpt for a program comprehension task\label{listing:example-program-comprehension}}]
    # Python code
    # Imports and pipeline definition...
    
    df = pd.read_csv('./data.csv')
    
    df.iloc[6:10, 0:3]
\end{lstlisting}

\begin{figure}[h]
  \centering
  \includegraphics[width=\linewidth]{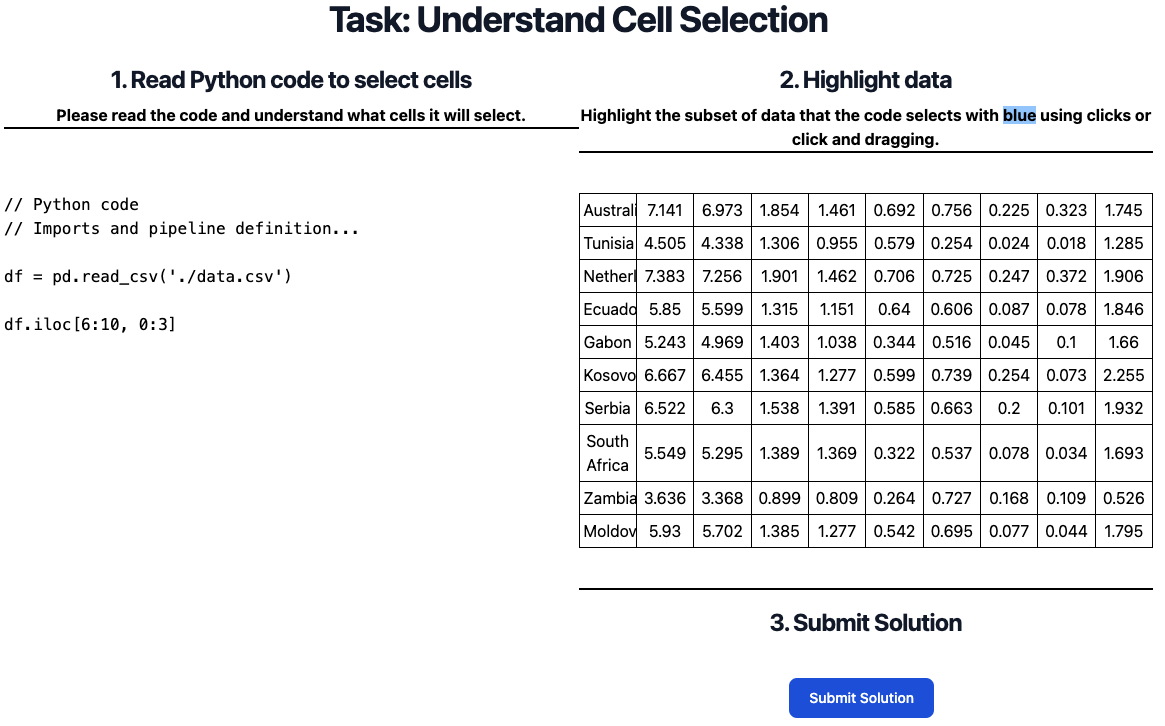}
  \caption{Program comprehension task in the experiment tool.}
  \label{fig:experiment-tool-pc}
\end{figure}

A complete task view for program comprehension is shown in \autoref{fig:experiment-tool-pc}.

\subsection{Instrumentation}
Before the experiment, participants were trained in Python and the DSL and their respective cell selection strategies during the university course.

Participants were introduced to the DSL and spreadsheets software cell selection syntax in two introduction lectures. During the semester, they completed five exercises with real-world open data using the DSL.

Based on their backgrounds, some participants already had some prior knowledge in Python and Pandas. In addition, they implemented a self-directed, real-world data science project in Python. The project included writing an automated data pipeline to download, clean and save open data sets before using the data to create a report.

I preparation for the experiment, we held a lecture on how to positionally select cells from 2D. The lecture included information about cell selection both in Python/Pandas using the iloc API used in the tasks as well as in the DSL using the same block used in the tasks.

The measurement instrument was a browser based environment that automatically tracked events for future analysis. Participants were asked not to use other programs or leave the experiment environment in any way. This explicitly included using websites to get outside help such as search engines or AI services. One experimenter monitored the screens of participants to make sure that they followed the directions.

At the start of the experiment, participants were allowed to open links to the official documentation of position-based indexing by Pandas, as well as the cell range documentation by the DSL. These links were provided by the experiment tool and the same for all participants.

\subsection{Data Collection Procedure}
The experiment was conducted in person in computer labs with identical equipment provided by the university. Due to the large size of the experiment, multiple sessions were conducted immediately following each other on one day and students were asked not to share the experiment setup with later groups.

After a brief introduction, the informed consent letter was handed out. The same letter was already shared by email beforehand. Participants were given the chance to ask questions or withdraw from the experiment without any negative consequences.

Then, one researcher followed a predefined experiment procedure document to present the experiment tool, the experiment flow, and two example tasks with pseudocode. Participants were asked to focus on correctness over speed if in doubt.

Data collection was done automatically using the experiment tool. For this, the tool automatically recorded events, their timing, and associated data like the submitted solution.

For each task, participants decided themselves when they considered a solution complete and submitted it. Between tasks, a success screen allowed them to pause before attempting the next task.

Participants were given 40 minutes to complete all tasks, with an announcement of time passing every 10 minutes.

\subsection{Analysis Procedure}
The analysis was completed using Python 3.11 with Pingouin 0.5.5 \cite{Vallat2018}.

First, experiment data was anonymized, and data integrity checks were performed, verifying that the tracked events are in the expected order and quantity. Another round of data integrity checks (such as verifying that no correctness value was outside the bound of 0 to 1) was executed after calculating the derived variables.

Second, from the timestamps of task start and end, the duration of a task was calculated in seconds. The overall time for a treatment is then taken as the average of both tasks completed using that treatment.

Correctness was calculated automatically by executing the code written by the participant and comparing the selected cells with the correct ones by building the Jaccard index, then averaging the correctness of both tasks. The more rigid structure of the DSL leads to slightly lower correctness values for code creation in the spreadsheet syntax for rare edge cases. We took note of this; however, with the correctness of the spreadsheet syntax being higher on average, this effect ultimately had no influence on the results of the hypothesis tests.

Few code creation submissions included syntax elements that were already part of the program snippet that was shown to participants, such as a trailing semicolon for the DSL or superfluous brackets for Pandas/Python. To reduce the amount of basic syntax errors that lead to correctness values of 0 with otherwise correct solutions, these syntax elements from the experiment tool were automatically removed for both Python and the DSL.

For the derived variables, outliers were marked using standard 1.5 interquartile range (IQR). Experiment runs with outliers in the variables under consideration were removed for the hypothesis test.
\section{Results}
\label{sec:results}
The experiment was performed with 100 participants, of which seven were removed due to an abnormal amount or order of events (mostly due to participants navigating back to previous tasks already completed during the experiment), leading to a final count of 93 valid experiment runs. Of these, 52 had been randomly assigned to the sequence BA, 41 to the sequence AB.

For the results, we chose to only consider experiment runs with no outliers in the variable under consideration, meaning individual hypothesis tests will have slightly different sample sizes depending on how many outliers were removed. Including experiment runs with outliers in the analysis slightly affects the individual values but does not change the results of the hypothesis tests.

We chose kernel-density plots to visualize the distributions of the variables because they provide a good overview of the distribution and make it easy to see non-normality \citep{Kitchenham2017-oi}. The plots are cut off at the extreme data points so that only existing data points are graphed.

The resulting distributions were analyzed for normality using the Shapiro-Wilk test \cite{Shapiro1965-fv}. Since all distributions were non-normal, we used the Wilcoxon signed-rank test for paired data for further hypothesis tests \cite{Wilcoxon1945-uf, Wohlin2012-ze}. We used the standard $\alpha = 0.05$ as a measure of statistical significance.

The detailed results of all hypothesis tests are shown in \autoref{tab:results-hypotheses-pc} and  \autoref{tab:results-hypotheses-cc}. We use common language effect size (CLES) as a more intuitive measure of effect size, first introduced by \cite{McGraw1992-ks}, but based on the generalization by \cite{Vargha2000-zj}, to discuss effect sizes.

The CLES describes the probability of a random value from one distribution to be larger than one from the other. Therefore, a value of $0.5$ is expected for no effect and larger deviations from that value with larger effects. We interpret CLES based on the guidelines in \cite{Vargha2000-zj} as either small, medium, or large (calculating $1 - CLES$ for values below $0.5$). For completeness, we additionally include effect sizes as matched pairs rank-biserial correlation (RBC) \cite{Kerby2014-xq}.

\subsection{Program Comprehension}

The results of the hypotheses tests for program comprehension are shown in \autoref{tab:results-hypotheses-pc}.

\begin{table}[h]
    \centering
    \caption{Wilcoxon signed-rank test results for program comprehension.}
    \label{tab:results-hypotheses-pc}
    \begin{tabular}{crrrrr}
    \toprule
    & n &  W-val & p-val & RBC & CLES \\
    \midrule
    $H_{1a}$ & 84 & 1459.0 & 0.146597 & -0.182633 & 0.464427\\
    $H_{1b}$ & 83 & 1230.0 & 0.018937 & 0.29432 & 0.541951\\
    \bottomrule
    \end{tabular}
\end{table}

We defined the null hypothesis for speed of program comprehension, $H_{0, 1a}$, as \textit{"Using spreadsheet-software cell selection syntax has no effect on the speed of bottom-up program comprehension compared to numeric indexing"}. With a p-value of $p \approx 0.147$ ($n = 84$), we have no reason to reject the null hypothesis and accept it as-is.

\begin{figure}[h]
  \centering
  \includegraphics[width=0.7\linewidth]{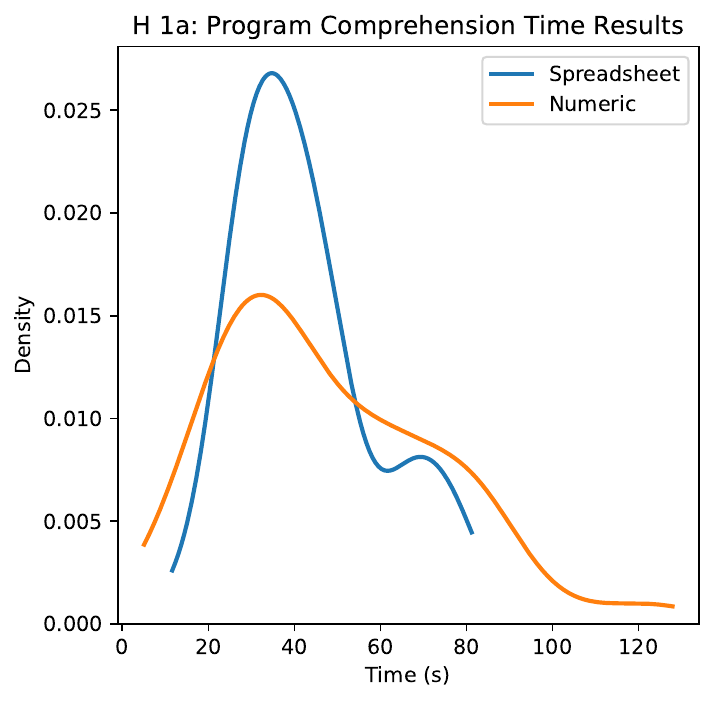}
  \caption{Kernel density plot for the results of $H_{1a}$, time, program comprehension}
  \label{fig:results-time-pc}
\end{figure}

From the distribution shown in \autoref{fig:results-time-pc}, it seems while participants using numeric syntax show more varied task completion times, both treatments have a similar peak submission time. Any existing difference is not large enough to be statistically relevant.

Regarding $H_{0, 1b}$, \textit{"Using spreadsheet-software cell selection syntax has no effect on the correctness of bottom-up program comprehension compared to numeric indexing"}, we reject the null hypothesis ($p \approx 0.019$, $n = 83$) and instead adopt the alternative hypothesis.

\begin{figure}[h]
  \centering
  \includegraphics[width=0.7\linewidth]{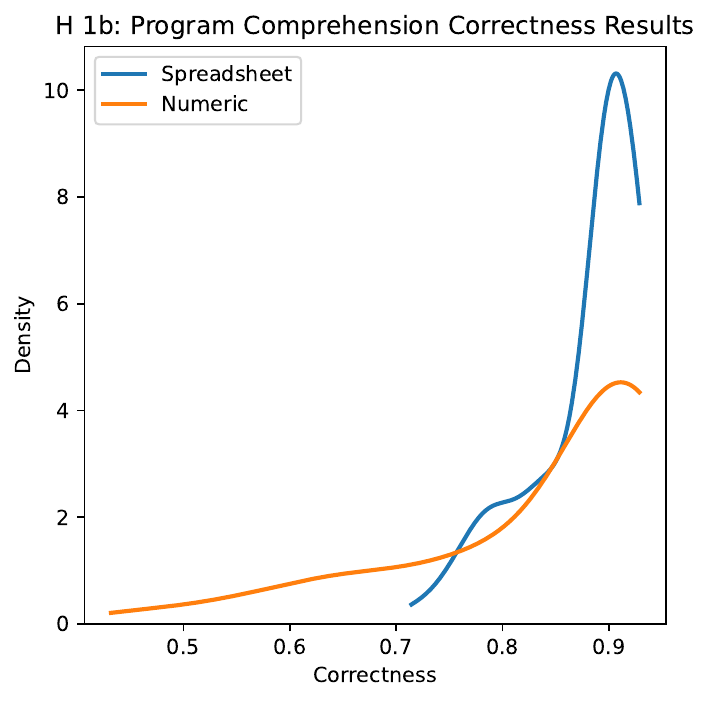}
  \caption{Kernel density plot for the results of $H_{1b}$, correctness, program comprehension}
  \label{fig:results-correctness-pc}
\end{figure}

From the distribution plotted in \autoref{fig:results-correctness-pc}, it is clear that participants understood cell selection significantly more correct when using spreadsheet syntax. This result is potentially very relevant to practice, as any data practitioner that wants to work with existing scripts first has to understand them and be confident that they are selecting the correct data.

However, with a CLES of $\approx 0.54$, the size of the effect is very small and has to be considered with that in mind. Further experiments should be conducted to verify that this effect does in fact exists.

\subsection{Code Creation}

The results of the hypotheses tests for code creation are shown in \autoref{tab:results-hypotheses-cc}.

\begin{table}[h]
    \centering
    \caption{Wilcoxon signed-rank test results for code creation.}
    \label{tab:results-hypotheses-cc}
    \begin{tabular}{crrrrr}
    \toprule
    & n &  W-val & p-val & RBC & CLES \\
    \midrule
    $H_{2a}$ & 84 & 163.0 & 4.775870e-13 & -0.908683 & 0.160289\\
    $H_{2b}$ & 93 & 463.5 & 0.000003 & 0.637324 & 0.658053\\
    \bottomrule
    \end{tabular}
\end{table}

\begin{figure}[h]
  \centering
  \includegraphics[width=0.7\linewidth]{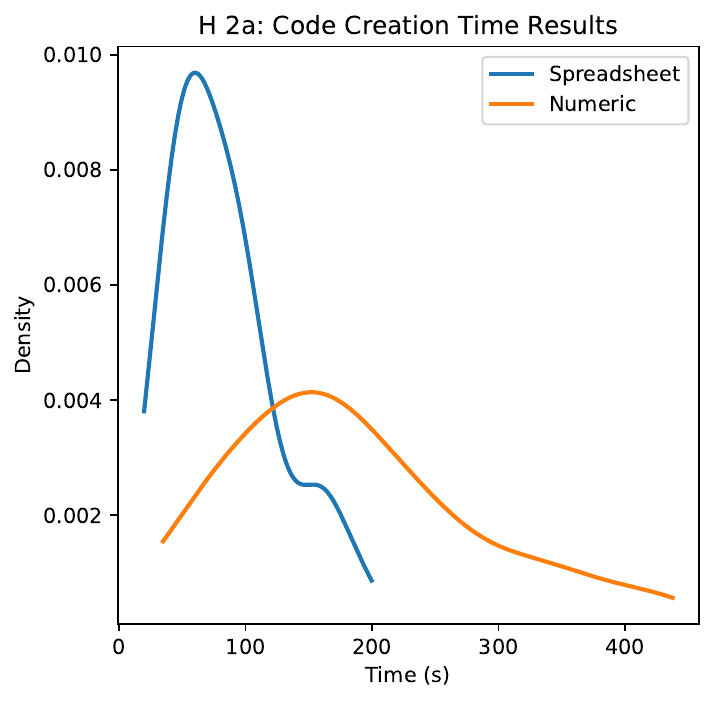}
  \caption{Kernel density plot for the results of $H_{2a}$, time, code creation}
  \label{fig:results-time-cc}
\end{figure}

The null hypothesis $H_{0, 2a}$, \textit{"Using spreadsheet-software cell selection syntax has no effect on the speed of code creation compared to numeric indexing"}, shows the largest effect. Participants are significantly faster to write code using the spreadsheet syntax ($p \approx 4.776e-13$, $n = 84$, see also \autoref{fig:results-time-cc}).

In addition, the effect size of $1 - 0.16 \approx 0.84 $ can be classified as large. Given how strong the effect is, using spreadsheet syntax to select cells from two-dimensional data has a clear impact on data practitioners and will allow them to complete their tasks faster.

Lastly, $H_{0, 2b}$, \textit{"Using spreadsheet-software cell selection syntax has no effect on the correctness of code creation compared to numeric indexing"} shows a statistically significant result as well ($p = 0.000003$, $n = 93$) with participants being more correct when using spreadsheet syntax to select cells instead of numeric indexing.

\begin{figure}[h]
  \centering
  \includegraphics[width=0.7\linewidth]{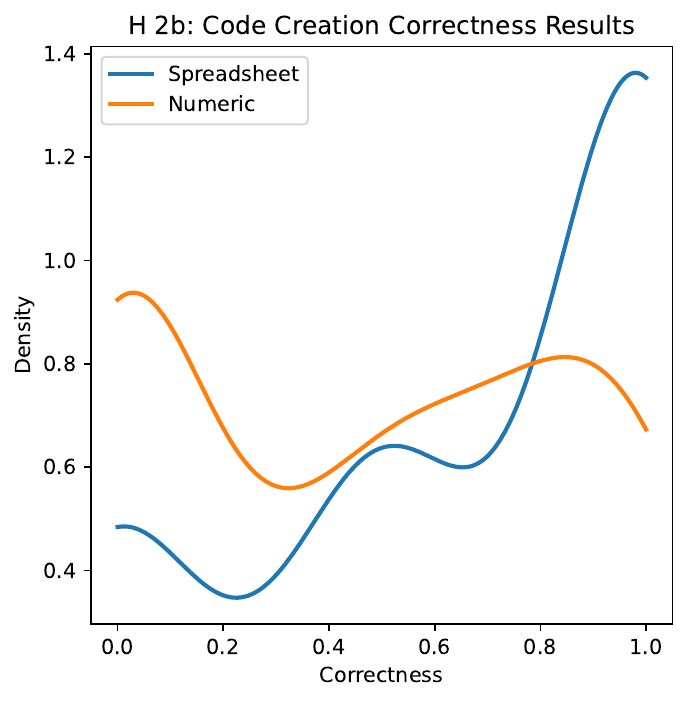}
  \caption{Kernel density plot for the results of $H_{2b}$, correctness, code creation}
  \label{fig:results-correctness-cc}
\end{figure}

From the kernel-density plot shown in \autoref{fig:results-correctness-cc} participants complete code creation tasks with much higher correctness using spreadsheet syntax than numeric syntax. Additionally, the rate of solutions with a correctness of 0 (mostly due to syntax errors) is dramatically lower when using the more simple spreadsheet syntax.

The effect size of $\approx 0.66$ is medium, with a high relevance to practice. Working with two-dimensional data is a common task for data engineers, and increased correctness will lead to lower amounts of bugs and further reduce the time needed to arrive at a correct solution.
\section{Discussion}
\label{sec:discussion}
In this chapter, we move beyond the quantitative results and discuss potential reasons for the effects that were measured in the experiment. Our goal is to provide additional insight from our extended engagement with the topic, the results and the participants; however, additional qualitative research should be done to rigorously describe causal explanations for the observed effects.

\subsection{On Program Comprehension}

With regards to program comprehension, using spreadsheet syntax resulted in higher correctness but had no statistically significant effect on the time needed.

Because the time needed to understand the cell selection syntax is similar, we assume participants largely did not try to verify their submission in depth by re-reading documentation and instead tried to answer from their intuitive understanding. It seems with this approach, both the spreadsheet style syntax as well as the numeric syntax are easy to read and did not create difficulties for the participants.

For correctness, the previous experience of participants very likely had an influence. During the coursework, many students pointed out that they often work with data sets using spreadsheet software like Microsoft Excel or Google Sheets. For many, these are standard tools used in data engineering to look at and edit small-scale 2D data sets. 
Similarly, practitioners from industry without a programming background often use spreadsheet software to work with data. Being able to reuse this previous experience can enable data practitioners to intuitively understand a selection syntax more correctly.

It also seems that choosing to start counting rows with one instead of sticking to the zero-indexing often found in GPLs means less confusion for users who do not have a programming background.

\subsection{On Code Creation}
When participants had to write code, they could complete the tasks in less time and with higher correctness when using the spreadsheet-style syntax compared to the numeric syntax.

A factor in the lower time needed to submit a solution might be the difference in the larger size of the documentation that is available for Pandas/Python. We encouraged the participants to, if in doubt, prioritize correctness over speed so it is conceivable that they verified their solution by re-reading the documentation.

On the other hand, this would also indicate that the spreadsheet syntax was more intuitive to understand and did require less double-checking with documentation.

It is important to point out that faster task completion is very likely at least partially predicated on the small scale of the dataset used. For larger data sets, especially when the spreadsheet syntax has to be extended and use two or more characters to refer to columns (e.g., the use of "AA" for the 27th column), the numeric syntax might be faster to use again.

Correctness was influenced by the comparatively larger amount of totally incorrect submissions with numeric indexing, which also includes syntax errors. The spreadsheet-style syntax has only one comparatively simple way to express a cell range. In contrast, the numeric syntax allows for some flexibility and is only one of many ways to select cells in Pandas. For example, some participants tried to refer to column names or used extraneous brackets or other not-allowed symbols, leading to syntax errors.

Aside from syntax errors, participants regularly made off-by-one errors using the numeric syntax, either incorrectly selecting more initial cells or missing a final row or column of cells. These errors seem to stem from a wrong intuition about zero-indexing selections in GPLs, as well as the resulting confusion about whether the final index is included or excluded from the selection. In contrast, the spreadsheet-style syntax seems to be more clear. One possible explanation is that the participants have previously selected data in the spreadsheet and visually seen the limits of their selection represented by the software.

\subsection{Potential Reasons for Improved Correctness}
A potential factor for improved correctness, both when reading code and when writing it, is the fact that spreadsheet syntax uses two different numbering schemes to refer to the two distinct indices in two-dimensional data.

Going back to the initial example, numeric syntax such as accessing the cell in the first row and second column with Pandas/Python using \lstinline{df.iloc[0, 1]} relies on the ordering of parameters. With this style, no additional context can be read from the syntax, and users have to rely purely on their knowledge about the correct position for, e.g., the index of the row they are trying to access.

In contrast, spreadsheet syntax such as the equivalent \lstinline{B1} refer to columns using characters and rows by numbers. This way, users can directly read which part of the syntax refers to which concept without having to rely on previous knowledge about the implementation. Instead, they have to be aware of the convention of referring to columns by characters, which they are due to their background using spreadsheet applications.

The different syntaxes are grounded in the two competing mental models when thinking about multi-dimensional data. The programmer's view (that finds its expression in the numeric syntax used by GPLs) is based on multi-dimensional arrays or matrices that are accessed along their axis. This mental model scales better to more than two dimensions as it does not assign inherent meaning to any axis.

The alternative mental model for thinking about two-dimensional data that is used by many data practitioners is viewing the data in a spreadsheet. While this view does not scale past two dimensions, it allows for the assignment of meaning to each axis and, therefore, custom representations for each. Additionally, using spreadsheet software with the permanent visual representation of row and column labels as numbers and characters reinforces an intuitive understanding of them among practitioners. 

\subsection{Practical Implications}
Overall, these results are an indication that a domain-specific syntax for cell selection should be considered when designing future languages for data engineering by data practitioners without a professional programming background.

The ability to contribute code faster hints at a lower technical barrier for users that have no background in software engineering, one of the major challenges to including subject-matter experts in collaborative data engineering \cite{Heltweg2023-ps}. Together with the reduced rate of errors, both when creating code and when understanding cell selections, these syntax adaptions could enable contributors with diverse backgrounds.

For language developers, knowing the target users' previous experience is most likely an important consideration. In this case, the insight that data practitioners often use spreadsheet programs to work with 2D data can directly lead to a more domain-appropriate syntax.
\section{Threats to Validity}
\label{sec:limitations}

We followed a thorough research process to conduct this study.
However, some potential threats to validity remain which we discuss according to the framework of threats to validity, as proposed in \cite{Wohlin2012-ze}.

\subsection{Threats to Conclusion Validity}

To make a valid conclusion, one must understand the correct relationships between the treatment and the results of an experiment.

Since the authors of this study are also creators oft he DSL used as a treatment in this study, searching for positive results might have introduced bias.
To mitigate this risk, we defined the hypotheses and the research design before the data collection. Further, we used standard research designs and statistical tests and report effect sizes and results regardless whether the results were statistically significant or not.

We employed a crossover experiment design to avoid the challenge of heterogeneity of students; in this manner, we measure differences in comparison to participants' average and not between participant groups  \cite{Vegas2016-dr}.

For the data collection, we strictly followed a previously designed experiment procedure document to reduce individual bias while guiding participants through the experiment. We automated large parts of the experiment in the form of an experiment tool that implements the treatment and took measurements without interaction by the researchers.

However, subconscious bias remains a potential threat to the validity of the conclusions. Therefore, we share the experiment tool for a thorough review and invite independent replication studies.

\subsection{Threats to Internal Validity}

To attribute observed effects solely to the treatment, it is crucial to control for any extraneous factors that might influence the outcome.

One such external factor is the quality of the tools and tasks of the experiment.
To ensure adequate quality, we tested the tool and the tasks in multiple sandbox tests with other researchers before using it during the experiment. We adjusted the tasks and the tool based on their feedback as suggested by Ko et al \cite{Ko2015-ti}.

The participation in the experiment was voluntary for the class of students.
The results might be biased by this selection if students expect that positive responses in regard to the DSL under study would positively influence their class grades.
To avoid this, we clearly communicated that the data was anonymized and emphasized that neither participation nor performance in the experiment influenced their grade.

Another external factor to discuss is carryover effects between treatments.
We addressed such potential learning effects by randomly assigning participants to different treatment orders and adding an initial task using pseudocode to allow familiarization with the tool and the task setup. 
Regardless of these measurements, we must recognize that carryover could still be an influencing factor on the results and aim for future replication with between-subject designs.

\subsection{Threats to Construct Validity}

To confirm that the measured variables accurately represent the intended theoretical constructs, it is essential to examine whether the operational definitions and instruments truly capture the underlying phenomena.

We clearly defined the dependent variables of the experiment and measured them programmatically. Further, we chose common measures for code comprehension and creation experiments: time and correctness \cite{Wyrich2023-ji}. However, while the Jaccard index we chose for correctness is a standard measurement, many competing definitions of correctness are possible.

Mono-method bias is a limitation for this study because we did only measure one variable for each construct. This presents a danger to insufficiently capture complex relationships, for example regarding program understanding. To strengthen the rigor of the results, additional experiments with more measurements would be needed. In future work, we plan to extend the current insights with more qualitative studies as well.

\subsection{Threats to External Validity}

To generalize the results of the experiment beyond its specific context, we need to carefully evaluate the applicability of the findings to other settings, populations, or times.

We chose students from master's degree programs in information systems, data science, and AI as participants for the experiment. For all drawn conclusions, it is important to contextualize them as representatives of a limited population, data practitioners that are not professional software engineers \cite{Falessi2018-so}.

However, we believe that those students are good proxies for a population of subject matter experts working with data in the industry and represent the variety of data practitioners.

Additional experiments with real subject-matter experts would be needed to validate whether students are a proxy, but we expect the results to generalize well in this limited domain.
We expect that the results do not generalize well to professional programmers with different previous backgrounds and more experiences with programming languages.
\section{Conclusions}
\label{sec:conclusions}
In conclusion, we conducted a large-scale, controlled experiment with student participants to find out if a domain-specific spreadsheet-style syntax had any effect on how well data practitioners select cells from 2D data, compared to the numeric syntax found in Pandas/Python.

In the experiment, participants completed tasks related to program comprehension by selecting a subset of cells as described by a program snippet. In addition, they had to complete a program with appropriate syntax to select the same cells that they were shown in a web-based tool.

With regards to program comprehension, we investigated if spreadsheet-style syntax had an effect on speed or correctness when reading cell selection code, compared to numeric indexing. Participants did understand the program more correctly when reading spreadsheet-style syntax but did not submit their solutions faster.

Similarly, for code creation, we measured time and correctness when completing program snippets with either spreadsheet-style syntax or numeric syntax. In these tasks, participants completed their tasks faster and more correctly when using spreadsheet-style syntax compared to numeric selection syntax.

From this data, we conclude that spreadsheet-style syntax can improve results for data practitioners when creating software artifacts for data engineering. Future language designers should consider the use of domain-specific syntax when targeting users who do not have a classical systems programmer background.

Concretely, the correctness of both reading and writing code was increased using spreadsheet syntax. This effect can improve the correctness of downstream data sets by reducing bugs in data pipeline code.

By using language syntax that is easier to use for practitioners and aligns more closely with their previous experience, technical barriers to participation by these users can be reduced. This in turn will allow more non-technical users, such as subject-matter experts, contribute to data engineering projects.

The implications for industry are important, with data engineering often consuming a large part of the costs for data science projects. Enabling contributors from a wider array of backgrounds to directly contribute with software artifacts can lower communication overhead and strain on professional software engineers.

While these results provide first quantitative indications, we can not draw clear causal explanations from them. To do so, additional qualitative research would be needed. By employing interviews or think-aloud protocols, the reasons for the effects of spreadsheet-style syntax should be explored in future work so that they can be used as guidelines for further language development.
\section*{Acknowledgments}
This research has been partially funded by the German Federal Ministry of Education and Research (BMBF) through grant 01IS17045 Software Campus 2.0 (Friedrich-Alexander-Universität Erlangen-Nürnberg) as part of the Software Campus project ’JValue-OCDE-Case1’. Responsibility for the content of this publication lies with the authors.
\section*{Data Availability Statement}
\label{sec:data-availability-statement}

The data generated and analyzed during the current study is available on Zenodo at \url{https://doi.org/10.5281/zenodo.15543965}.

\bibliographystyle{plain}
\bibliography{arxiv}

\end{document}